\documentclass{iau}
\usepackage{graphicx}

\newcommand{\msun}{$\mathrm{M_{\odot}}$}

\title[Evolution of Rossby number] %% give here short title %%
{Theoretical evolution of Rossby number for solar analog stars}

\author[M. Castro et al.]   %% give here short author list %%
{Matthieu Castro,
 Tharc\'isyo Duarte
 \and Jos\'e Dias do Nascimento Jr.}

\affiliation{Departamento de F\'isica Te\'orica e Experimental (DFTE), Universidade Federal do Rio Grande do Norte (UFRN), Campus Universit\'ario Lagoa Nova, CEP: 59078-970, Natal (RN), Brazil \\ email: {\tt mcastro@dfte.ufrn.br}}

\pubyear{2013}
\volume{302}  %% insert here IAU Symposium No.
\pagerange{xxx--xxx}
\jname{Magnetic Fields Throughout Stellar Evolution}
\editors{A.C. Editor, B.D. Editor \& C.E. Editor, eds.}
\begin{document}

\maketitle

\begin{abstract}
Magnetic fields of late-type stars are presumably generated by a dynamo mechanism at the interface layer between the radiative interior and the outer convective zone. The Rossby number, which is related to the dynamo process, shows an observational correlation with activity. It represents the ratio between the rotation period of the star and the local convective turnover time. The former is well determined from observations but the latter is estimated by an empirical iterated function depending on the color index $(B-V)$ and the mixing-length parameter. We computed the theoretical Rossby number of stellar models with the TGEC code, and analyze its evolution with time during the main sequence. We estimated a function for the local convective turnover time corresponding to a mixing-length parameter inferred from a solar model, and compare our results to the estimated Rossby number of 33 solar analogs and twins, observed with the spectropolarimeters ESPaDOnS@CFHT and Narval@LBT.
\keywords{solar analogs, magnetic field, Rossby number}
\end{abstract}

\firstsection % if your document starts with a section,

\section{Evolutionary models}
\label{models}

Stellar evolution calculations were computed with the Toulouse-Geneva stellar evolution code TGEC. Details of the input physics used in our models can be found in \cite{donascimento13}. The evolution of the angular momentum is calculed with the \cite{kawaler88} law, and for the initial rotation rates, we adopted the relation (3) in \cite{landin10}. The calibration method of the models is based on the \cite{richard04} prescription: a solar model is calibrated to match the observed solar radius and luminosity at the solar age. The parameter $K$ of the angular momentum evolution law is adjusted to give the solar rotation period ($P_{\mathrm{rot}} = 27.1$ d) at the solar age. Evolutionary models of masses 0.81, 0.85, 0.90, 0.95, 1.00, 1.05, and 1.10 \msun\ were calculated from the ZAMS to the top of the RGB. The input parameters for the other masses are the same as for the 1.00 \msun\ model. 

% Figure \ref{Prot} shows the evolution of rotation period of our models as a function of effective temperature.
% 
% \begin{figure}[h]
% % \vspace*{-2.0 cm}
% \begin{center}
%  \includegraphics[width=2.2in]{castro.matthieu_fig1.eps} 
% % \vspace*{-1.0 cm}
%  \caption{Rotation period as a function of effective temperature for evolutionary models of different masses; filled black circles are observed stars by NARVAL@TBL and ESPaDOnS@CFHT. The red circle is the solar twin HIP 55459 and the position of the Sun is indicated.}
%    \label{Prot}
% \end{center}
% \end{figure}

\section{Stars sample}
\label{sample}

To compare with our models, we use a sample of 33 sun-like stars observed with the spectropolarimeters ESPaDOnS at the Canada-France-Hawaii Telescope (Mauna Kea, USA) and NARVAL at the T\'elescope Bernard Lyot (Pic du Midi, France).
%  Thanks to this new generation of spectropolarimeters, we are able to directly measure the magnetic field of a low-mass star from the Zeeman signature in the spectrum. For low-activity solar-type stars, we expect a maximum amplitude of circularly polarized Zeeman signatures of the order of $10^{-4} I_{\mathrm{C}}$, where $I_{\mathrm{C}}$ denotes the continuum level. A signal-to-noise ratio (S/N) as high as 50,000 is therefore needed for the detection of Stokes V signatures with a confidence level of $5\sigma$. To reach this high polarimetric accuracy, we collected stellar spectra with S/N between 1,000 and 1,500 per 2.6 km/s velocity bin. The technique of Least-Square-Deconvolution (LSD, \cite[Donati et al. 1997]{donati97}), by simultaneously extracting the information contained in all 5,000 photospheric lines of the echelogram (for a line-list matching a solar atmospheric model), decreases the noise level by a factor of about 35. 
Longitudinal magnetic field of the stars were determined with the technique of Least-Square-Deconvolution (LSD, \cite[Donati et al. 1997]{donati97}). Rotation periods of the stars, determined from chromospheric activity, were found in the literature (\cite[Noyes et al. 1984]{noyes84}, \cite[Wright et al. 2004]{wright94}, \cite[Lovis et al. 2011]{lovis11}).

\section{Rossby number}
\label{rossby}

\begin{figure}[h]
 \vspace*{-0.35 cm}
\begin{center}
  \includegraphics[width=1.74in]{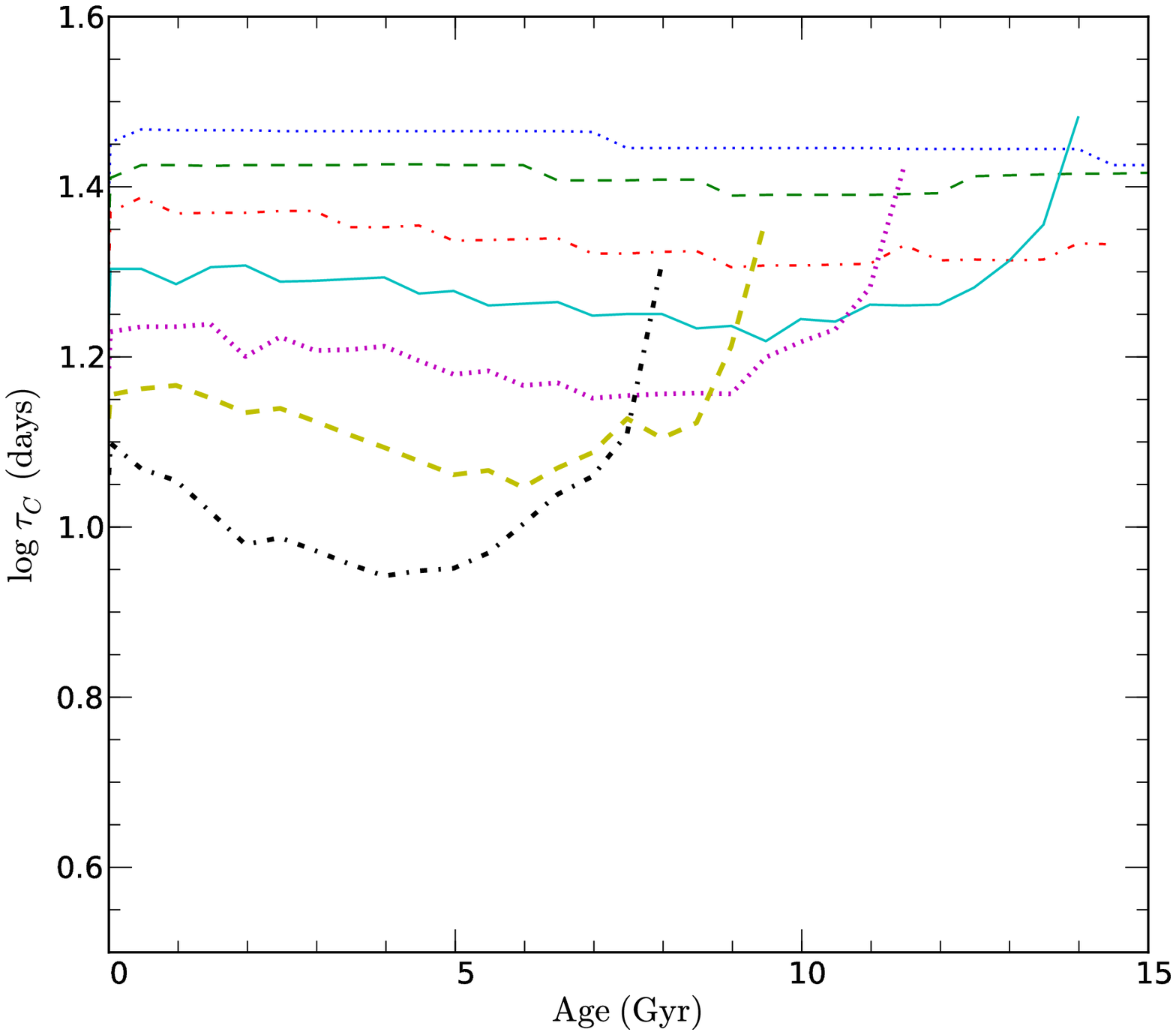}
  \qquad
  \includegraphics[width=1.7in]{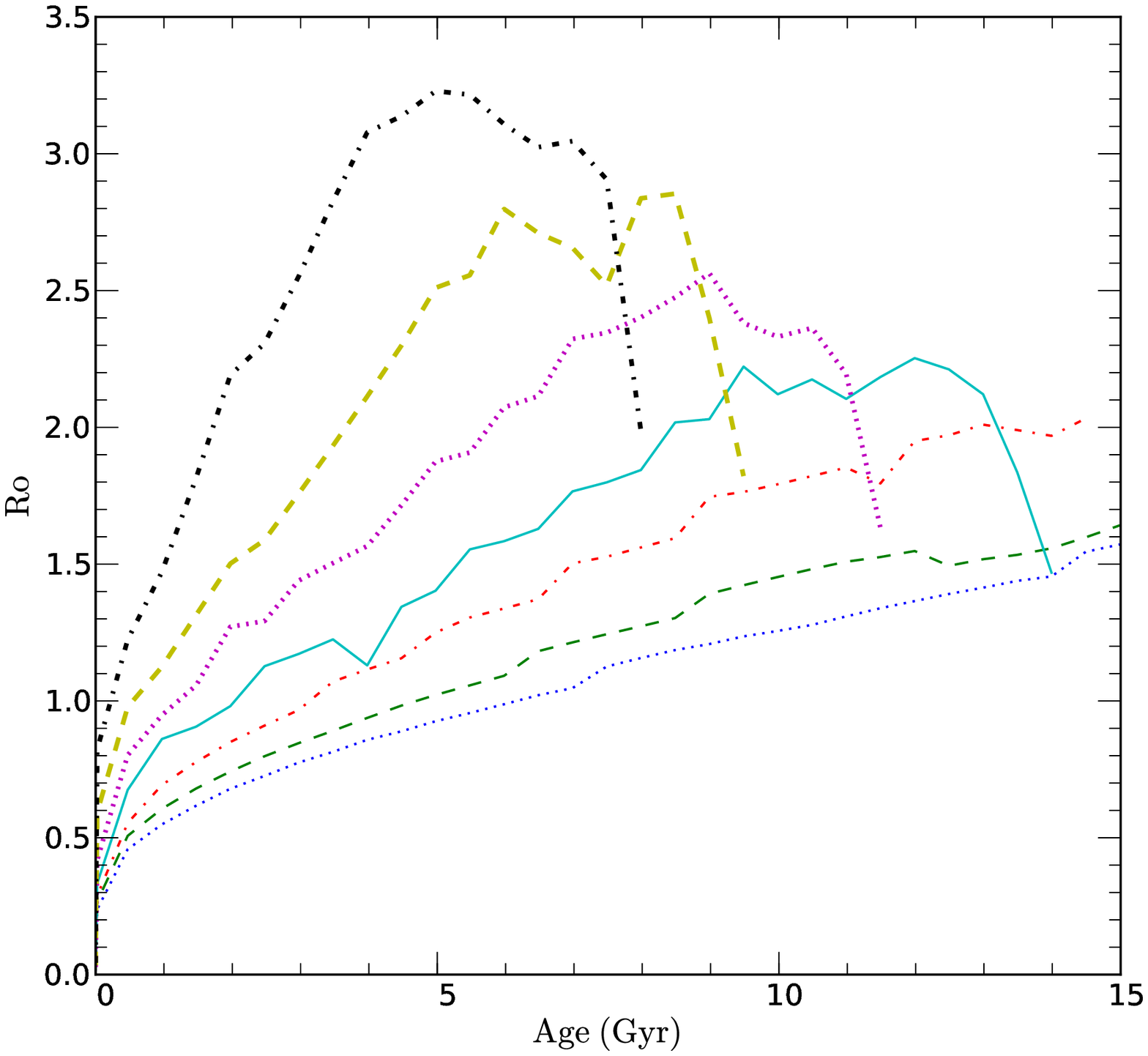} 
\vspace*{-0.2 cm}
 \caption{$\log \tau_{\mathrm{C}}$ (\textit{left}) and Rossby number $Ro$ (\textit{right}) as a function of age for our models of 0.81 (dotted), 0.85 (dashed), 0.90 (dot-dashed), 0.95 (continuous), 1.00 (thick dotted), 1.05 (thick dashed), and 1.10 (thick dot-dashed) \msun.}
   \label{logtauc&Ro_age}
\end{center}
\end{figure}

For all the models, we calculated the Rossby number $Ro = P_{\mathrm{rot}}/\tau_{\mathrm{c}}$ where $P_{\mathrm{rot}}$ is the rotation period, and $\tau_{\mathrm{c}}$ the local convective turnover time, calculated at a distance of one pressure height scale $H_{\mathrm{P}}$ above the base of the convective zone. Its value is computed through the equation $\tau_{\mathrm{c}} = \alpha H_{\mathrm{P}}/v$, where $v$ is the convective velocity. In Fig. \ref{logtauc&Ro_age}, we plot the evolution of $\log \tau_{\mathrm{c}}$ (left panel) and $Ro$ (right panel) as a function of the stellar age.

\begin{figure}[h]
\vspace*{-0.35 cm}
\begin{center}
  \includegraphics[width=1.74in]{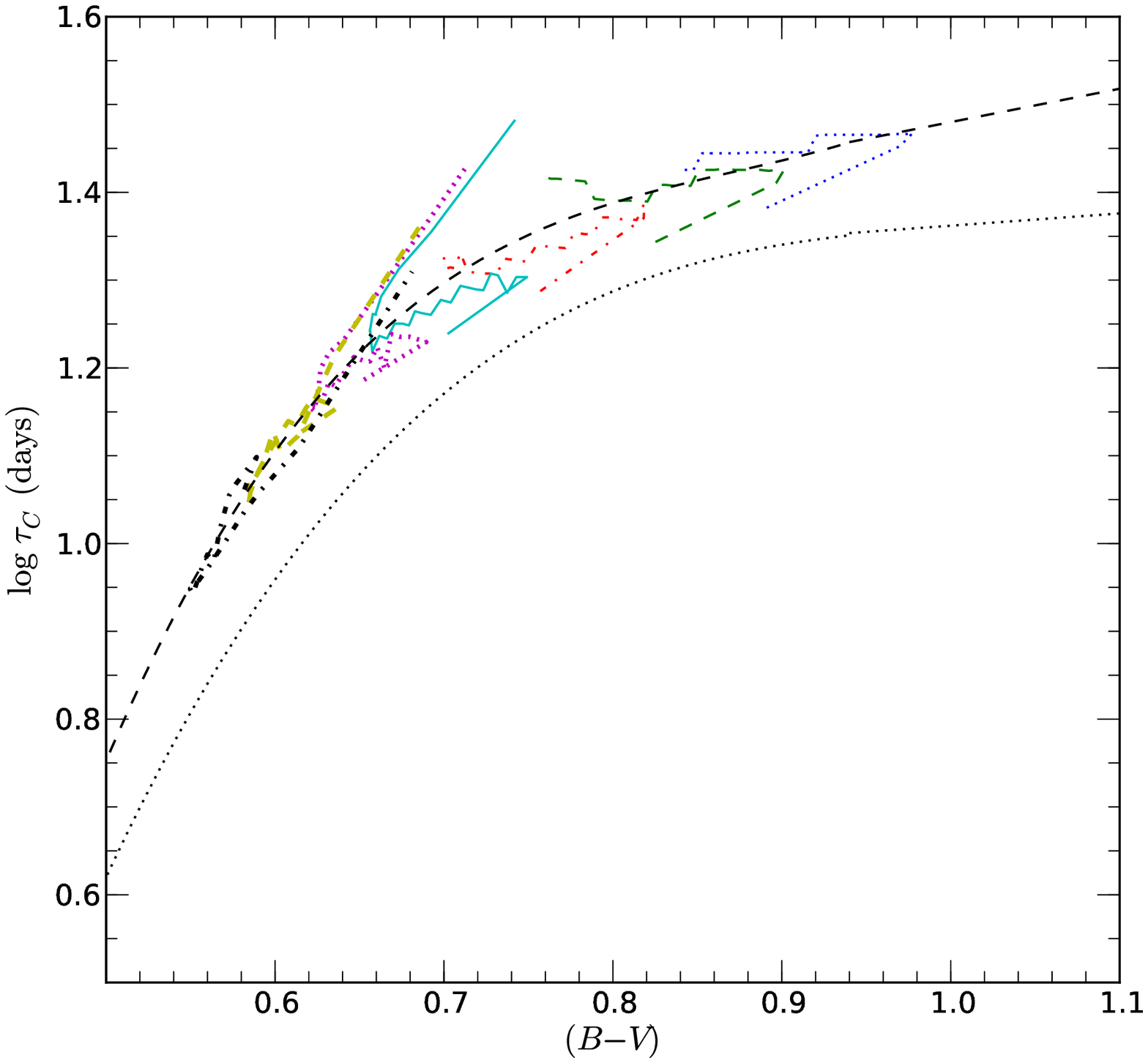}
  \qquad
  \includegraphics[width=1.7in]{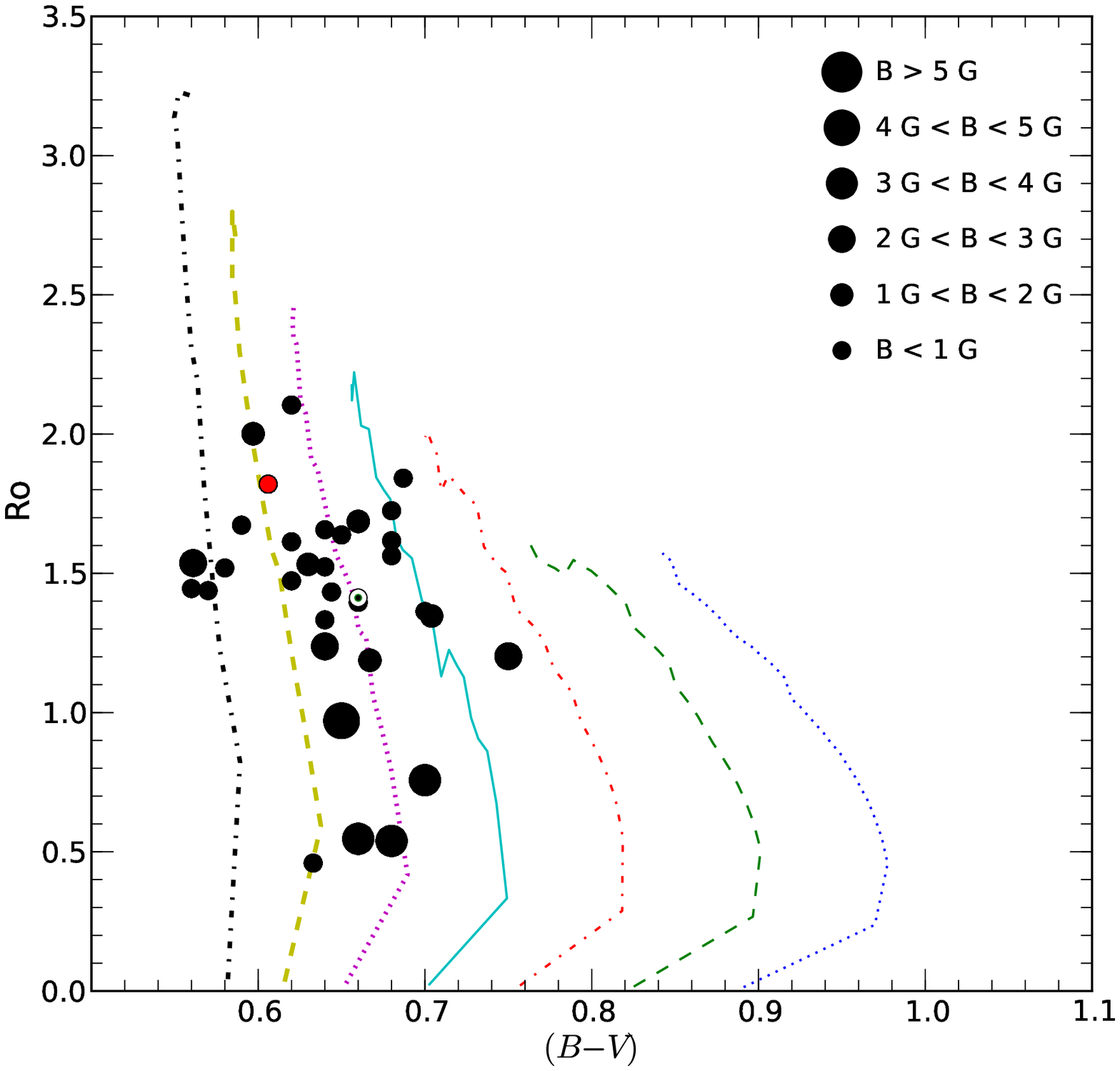} 
 \vspace*{-0.2 cm}
 \caption{\textit{Left}: $\log \tau_{\mathrm{c}}$ as a function of $(B-V)$. The dashed line is the polynomial regression of the models tracks. The dotted line is the polynomial regression found by \cite{noyes84}. \textit{Right}: Rossby number $Ro$ as a function of $(B-V)$ for our models limited to the main sequence. Filled black circles are observed stars, the size depends on the strength of the magnetic field.}% The red circle is the solar twin HIP 55459 and the position of the Sun is indicated.}
   \label{logtauc&Ro_BV}
\end{center}
\end{figure}

In the left panel of Fig. \ref{logtauc&Ro_BV}, we show the $\log \tau_{\mathrm{c}}$ of our models as a function of $(B-V)$. A cubic fit through these curves (dashed line) defines an empirical function $\log \tau_{\mathrm{C}}(B-V)$, given by: $\log \tau_{\mathrm{c}} = -5.468 + 22.810(B-V) - 25.637(B-V)^2 + 9.796(B-V)^3$. From this equation, we determine the $\tau_{\mathrm{c}}$ of the stars of our sample, and then a Rossby number $Ro$. In the right panel of Fig. \ref{logtauc&Ro_BV} we plot the Rossby number as a function of $(B-V)$ for our models and for the stars of our sample, determined from the observed rotation period and the convective turnover time inferred from the above equation.% The size of the circles depends on the strength of the observed magnetic field.

\begin{thebibliography}{}

\bibitem[do Nascimento et al. (2013)]{donascimento13}
{do Nascimento, J.-D., da Costa, J. S. \& Castro, M.} 2013,
\textit{A\&A}, 548, L1

\bibitem[Donati et al. (1997)]{donati97}
{Donati, J.-F., Semel, M., Carter, B. D. \etal\ } 1997,
% {Donati, J.-F., Semel, M., Carter, B. D., Rees, D. E. \& Collier Cameron, A.} 1997,
\textit{MNRAS}, 291, 658

% \bibitem[Hui-Bon-Hoa (2008)]{huibonhoa08}
% {Hui-Bon-Hoa, A.} 2008, 
% \textit{Ap\&SS}, 316, 55

% \bibitem[Kawaler (1987)]{kawaler87}
% {Kawaler, S. D.} 1987, 
% \textit{PASP}, 99, 1322

\bibitem[Kawaler (1988)]{kawaler88}
{Kawaler, S. D.} 1988, 
\textit{ApJ}, 333, 236

\bibitem[Landin et al. (2010)]{landin10}
{Landin, N. R., Mendes, L. T. S. \& Vaz, L. P. R.} 2010,
\textit{A\&A}, 510, A46

\bibitem[Lovis et al. (2011)]{lovis11} 
{Lovis, C., Dumusque, X., Santos, N. C.  \etal\ } 2011,
% {Lovis, C., Dumusque, X., Santos, N. C.  Bouchy, F., Mayor, M., Pepe, F., Queloz, D., S\'egransan, D. \& Udry, S.} 2011,
eprint arXiv:1107.5325

\bibitem[Noyes et al. (1984)]{noyes84} 
{Noyes, R. W., Hartmann, S. W., Baliunas, S. \etal\ } 1984,
% {Noyes, R. W., Hartmann, S. W., Baliunas, S., Duncan, D. K. \& Vaughan, A. H.} 1984,
\textit{ApJ}, 279, 763

% \bibitem[Paquette et al. (1986)]{paquette86}  
% {Paquette, C., Pelletier, C., Fontaine, G. \& Michaud, G.} 1986, 
% \textit{ApJS}, 61, 177

% \bibitem[Pinsonneault et al. (1989)]{pinsonneault89}
% {Pinsonneault, M. H., Kawaler, S. D., Sofia, S. \& Demarque, P.} 1989, 
% \textit{ApJ}, 338, 424

\bibitem[Richard et al. (2004)]{richard04}
{Richard, O., Théado, S. \& Vauclair, S.} 2004,
\textit{SoPh}, 220, 243

% \bibitem[Talon \& Zahn (1997)]{talon&zahn97}
% {Talon, S. \& Zahn, J.-P.} 1997, 
% \textit{A\&A}, 317, 749

\bibitem[Wright et al. (2004)]{wright94}
{Wright, J. T., Marcy, G. W., Paul Butler, R. \& Vogt, S. S.} 2004,
\textit{ApJS}, 152, 261

% \bibitem[Zahn (1992)]{zahn92} 
% {Zahn, J.-P.} 1992, 
% \textit{A\&A}, 265, 115


\end{thebibliography}
\end{document}